\documentclass[conference]{IEEEtran}
\IEEEoverridecommandlockouts
\usepackage{cite}
\usepackage{amsmath,amssymb,amsfonts}
\usepackage{algorithmic}
\usepackage{graphicx}
\usepackage{epstopdf}
\usepackage{textcomp}
\usepackage{xcolor}
\usepackage{footnote}
\usepackage{booktabs}
\usepackage[colorlinks,linkcolor=black]{hyperref}
\hypersetup{
    colorlinks=true,
    linkcolor=black,
    urlcolor=black,
    citecolor=black,
    anchorcolor=black
}

\newcommand{\ie}{\textit{i}.\textit{e}.} 
\newcommand{\eg}{\textit{e}.\textit{g}.} 

\def\BibTeX{{\rm B\kern-.05em{\sc i\kern-.025em b}\kern-.08em
    T\kern-.1667em\lower.7ex\hbox{E}\kern-.125emX}}
\begin{document}

\title{PianoBART: Symbolic Piano Music Generation and Understanding with Large-Scale Pre-Training 
}


\author{
    \IEEEauthorblockN{Xiao Liang, Zijian Zhao, Weichao Zeng, Yutong He, Fupeng He, Yiyi Wang, Chengying Gao$^{*}$\thanks{$^*$The corresponding author is Chengying Gao. }}
    \IEEEauthorblockA{School of Computer Science and Engineering, Sun Yat-sen University, Guangzhou, China}
    \IEEEauthorblockA{\{liangx66, zhaozj28, zengwch6, heyt53, hefp6, wangyy339\}@mail2.sysu.edu.cn, mcsgcy@mail.sysu.edu.cn}
}

\maketitle

\begin{abstract}
Learning musical structures and composition patterns is necessary for both music generation and understanding, but current methods do not make uniform use of learned features to generate and comprehend music simultaneously.
In this paper, we propose PianoBART, a pre-trained model that uses BART for both symbolic piano music generation and understanding.
We devise a multi-level object selection strategy for different pre-training tasks of PianoBART, which can prevent information leakage or loss and enhance learning ability.
The musical semantics captured in pre-training are fine-tuned for music generation and understanding tasks. 
Experiments demonstrate that PianoBART efficiently learns musical patterns and achieves outstanding performance in generating high-quality coherent pieces and comprehending music.
Our code and supplementary material are available at \href{https://github.com/RS2002/PianoBart}{https://github.com/RS2002/PianoBart}.
\end{abstract}

\begin{IEEEkeywords}
Automatic Music Generation, Music Understanding, Symbolic Piano Music, Bidirectional and Auto-Regressive Transformers (BART)
\end{IEEEkeywords}

\section{Introduction}
Music generation and understanding are interrelated topics in the music community.
Understanding the melody, rhythm, and structure \cite{MusicBERT, MidiBERT} greatly benefits effective music generation.
Music generation \cite{MusicTransformer, StyleMusicGeneration} is also helpful in investigating how well machines understand musical structure and composition patterns.
Hence there is great and urgent interest in exploring automatic music generation and understanding.

Given the sequential similarity between text and symbolic music, language-based methods have been applied to symbolic music generation and understanding \cite{MusicTransformer, Pop_music_transformer, StyleMusicGeneration, MusicBERT, MidiBERT, InformationLeakage}.
However, directly using language-based methods is challenging due to the inherent differences between text and music.
First, music has a more complex semantic and hierarchical structure than natural language, which means that symbolic music requires much longer sequences to represent.
Second, symbolic music involves various musical elements like melody, rhythm, and harmony. 
Nevertheless, the lack of sufficient specialized labeled data hinders representation learning.

Although recent works on music generation have proposed some encoding methods to represent symbolic music \cite{Pop_music_transformer, CP}, these encodings produce very long sequences up to thousands or more tokens for a full song of several minutes.
To address this, we adopt the compact Octuple encoding \cite{MusicBERT}. 
It can efficiently and comprehensively represent music and greatly reduce the sequence length, thus supporting long-term music generation and full-song-level understanding.

In many fields including Music Information Retrieval (MIR), there is often insufficient labeled data due to factors like expensive cost and copyright issues. 
To overcome the scarcity of labeled data, pre-training on available unlabeled data has become the most promising method \cite{CSIBERT, CV_mask, EL_pretrain}, as it enables models to learn common features of data structure. 
In MIR, there have been several works \cite{MusicBERT, MidiBERT, InformationLeakage} using the pre-training models in NLP such as BERT \cite{BERT}.
However, the problem of information leakage or loss is caused by these models selecting too few or too many objects for pre-training transformations, which degrades the ability to capture underlying musical patterns. 
To prevent information leakage or loss, based on the octuple encoding and pre-training tasks, we propose a multi-level object selection strategy including a designed n-bar level method.
The method can dynamically determine the selection range of pre-training objects and is applicable to different pre-training transformations.

Both music generation and music understanding are based on learning musical structure and semantics, while existing methods address these tasks separately, preventing the reuse of learned features in one task for the other. 
To this end, we propose PianoBART, a novel BART-based pre-trained model that addresses both symbolic piano music generation and understanding in a unified framework. 
Unlike BERT-based pre-trained models \cite{MusicBERT, MidiBERT, InformationLeakage}, PianoBART uses an encoder-decoder structure that allows it to be applied for sequence-to-sequence task, thus handling both generation and comprehension. 
In addition, PianoBART includes more transformations than BERT to enhance pre-training effectively. 
More importantly, for each transformation, we design corresponding object selection methods to largely avoid information leakage or loss. 
PianoBART is pre-trained in a self-supervised way to overcome the lack of labeled data.
Experiments show that PianoBART successfully captures music domain knowledge and excels in both music generation and comprehension.

\begin{figure*}[h]\centering
  \includegraphics[width=\textwidth]{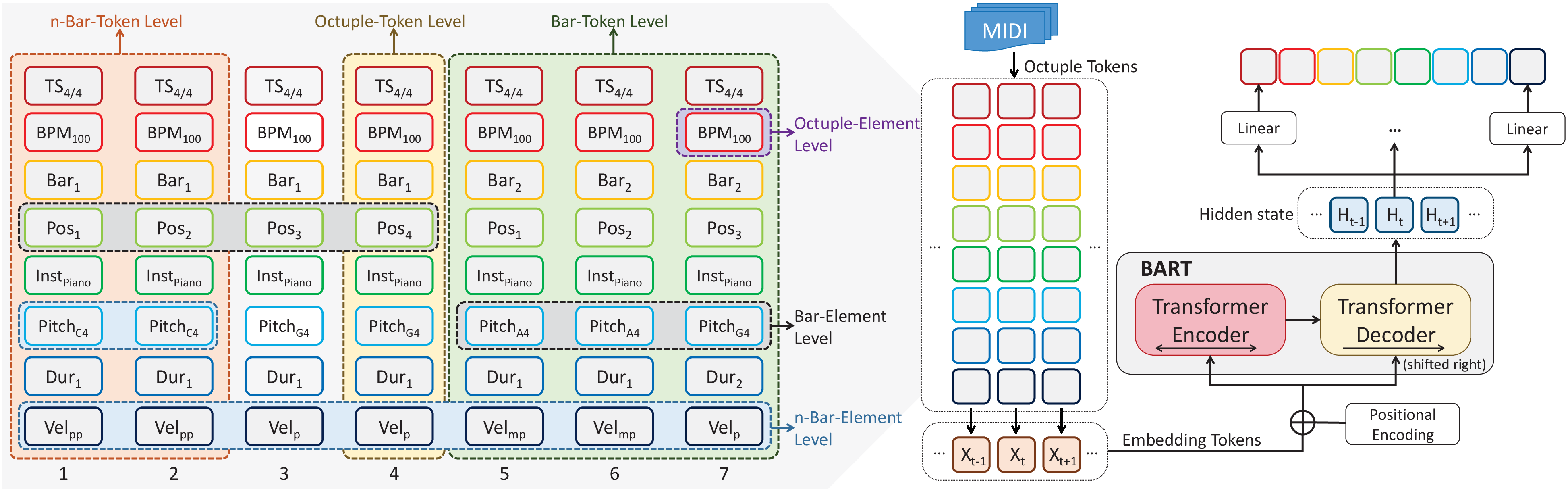}
  \caption{  
Overview of the proposed PianoBART framework (right) and the designed multi-level object selection strategy (left).
} 
  \label{fig:architecture}
\end{figure*}

In summary, the main contributions of this work are:

(1) We propose PianoBART, the first large-scale pre-trained model that uses BART for both symbolic piano music generation and understanding.

(2) A novel multi-level object selection strategy for pre-training is designed to avoid information leakage and loss while improving the quality of downstream tasks.

(3) Experiments show that PianoBART achieves excellent performance on realistic and coherent music generation and a number of music understanding tasks.


\section{Related Work}\label{Related Work}

\subsection{Symbolic Music Generation}

Significant progress has been made in automatic symbolic music generation \cite{MusicTransformer, StyleMusicGeneration, Pop_music_transformer}.
Inspired by the similarity between text and symbolic music, sequence models like attention-based Transformer \cite{Transformer} have been increasingly applied to capture the long-term dependency of music and produce coherent music samples \cite{Pop_music_transformer, CP}.
However, existing symbolic music encoding \cite{Pop_music_transformer, Popmag} produces too lengthy sequences, making the Transformer computationally difficult.
In this paper, we introduce Octuple representation \cite{MusicBERT} into music generation, effectively reducing the sequence length.
Besides, this paper introduces the BART pre-training model \cite{BART} and proposes PianoBART for symbolic piano music generation.

\subsection{Symbolic Music Understanding}
In the symbolic music domain, it's difficult and time-consuming to obtain professional labels, hence existing labeled datasets remain small size \cite{POP909, ASAP, EMOPIA}.
To overcome the lack of labeled data, recent works use pre-trained language models like BERT \cite{BERT, Roberta} to learn the long-term musical structure in an unsupervised way \cite{MusicBERT, MidiBERT, InformationLeakage}.
However, due to the repetitive characteristics of music, simply using pre-training methods in NLP (e.g., mask language model \cite{BERT}) for symbolic music may lead to information leakage or loss \cite{MusicBERT}, affecting the performance of downstream tasks.
To address this, we design a multi-level object selection strategy for pre-training that is able to enhance the model's robustness.

\section{Approach}\label{Approach}
\subsection{PianoBART Framework} \label{PianoBART Framework}
The proposed PianoBART is a BART-based model for piano music generation and understanding.
As shown in Fig.~\ref{fig:architecture}, PianoBART adopts the standard Transformer encoder and decoder \cite{Transformer} architecture as the backbone.
The encoder bidirectionally encodes sequences of symbolic music tokens through multi-head self-attention.
The decoder autoregressively generates outputs from left to right, which is suitable for sequence generation tasks.

We employ the Octuple representation \cite{MusicBERT} to encode symbolic music.
An example is demonstrated in Fig.~\ref{fig:architecture}, where the input MIDI is converted to a sequence of octuple tokens.
Each octuple token corresponds to a note and contains 8 musical elements, including time signature (TS), tempo (BPM), bar, position, instrument, pitch, duration, and velocity.
The embeddings of the 8 elements in each token are concatenated together and then linear projected to the embedding token, which is fed to the BART.
As for the hidden state produced by the Transformer decoder, octuple elements in each token can be predicted simultaneously with different linear layers.

PianoBART is pre-trained by (1) corrupting octuple token sequences, and (2) learning a model to reconstruct the original sequences \cite{BART}.
More diverse transformations than BERT that destroy the sequence structure are employed to enhance the model's ability to learn the musical pattern.
We design a multi-level object selection strategy based on the pre-training transformations and the employed Octuple encoding. 
This strategy can effectively prevent the information leakage and loss that may occur during the pre-training.
PianoBART is fine-tuned on a variety of downstream tasks in music generation and understanding, which will be described in experiments.

\subsection{Multi-level Object Selection Strategy} \label{Multi-level Object Selection Strategy}
PianoBART is trained in a self-supervised way that maps the corrupted octuple token sequences into the original ones.
The initial operation is to select specific objects and apply certain transformations to corrupt the original structure. 
Then the model is trained to reconstruct the modified objects.

To choose the objects to be corrupted, we design a multi-level object selection strategy from two dimensions of the octuple encoding, \ie, attributes and time span.
For simplicity, we refer to each octuple object as a ``token'' and each attribute within an octuple object as an ``element''.
In terms of attributes, we consider the Element Level and the Token Level, with element and token as a single entity, respectively.
In terms of time span, we consider the Octuple Level, the Bar Level, and the n-Bar Level, each covering a different period.
The combination of attributes and time span leads to 6 selection methods, as shown in Table \ref{tab:table1}.

\noindent
\textbf{Octuple Level.}
Figure~\ref{fig:architecture} shows an example of the Octuple Level selection method, which is inspired by the naive selection method in the mask language modeling (MLM) of BERT \cite{BERT}. 
In the Octuple-Element Level and the Octuple-Token Level, we randomly choose independent elements or tokens as target objects, respectively.

However, the octuple level method may cause information leakage.
In specific, music is repetitive, some musical attributes (\eg, bar, position, and pitch) may be identical in successive segments. 
Considering the case of simply masking a single note, the missing attributes can be easily inferred by directly replicating neighboring notes.
The model can therefore achieve relatively high accuracy without learning the music context.
However, the underlying musical structures and patterns cannot be fully captured, which limits the performance of downstream tasks.

\noindent
\textbf{Bar Level.} 
To address the information leakage problem, the Bar-Element Level method is proposed by \cite{MusicBERT}, where elements of the same type in the same bar are regarded as a unit and selected simultaneously.
Moreover, we further present the Bar-Token Level method, which chooses all the complete tokens within the same bar at the same time.

However, in musical compositions, some elements do not strictly repeat within a whole bar.
For example, as shown in Fig.~\ref{fig:architecture}, the ``$Pitch_{A4}$'' only repeats for half a bar. 
Although the information leakage is mitigated at the bar level, it's more likely to cause information loss, when a whole bar that may contain dozens of notes or tokens is masked \cite{InformationLeakage}.

\noindent
\textbf{n-Bar Level.} 
To enhance the model's generalization capacity, we further design a novel n-Bar Level Method, where the time span $n$ is randomly chosen and defined as follows:

\begin{equation}
    n=\inf\{n: \sum_{i=p}^{p+n} dur(T_i) \geq \frac{m}{64}\}.
    \label{equ:nBar}
\end{equation}
where $p$ is the number of an initially selected token $T_p$, 
$T_i$ is the $i^{th}$ token in the octuple sequence, 
$dur()$ represents the duration of a token, 
and $m$ is a random integer in $[1, 128]$.
The n-Bar-Element Level or the n-Bar-Token Level is to select $n$ consecutive elements or tokens at a time, respectively.
For example, assuming there are 4 quarter-notes in a measure, only the first note is selected ($n=1$) if $m\in [1, 16]$, the first two notes are selected ($n=2$) if $m\in [17, 32]$, and so on.

According to Equation~\eqref{equ:nBar}, the minimum selected time span is a hemidemisemiquaver ($1/64$) and the maximum is two whole notes (2 bars in 4/4 time signature), which is the common range of duration for a note.
Compared to the Bar Level method (equivalent to the 1-Bar-Level), the selection range of n-Bar Level is optional, not limited to a single bar, which is more flexible.
This way can effectively prevent information loss and avoid information leakage.
Furthermore, the dynamic selection range helps the model to learn more structural relations in music like intra-bar and inter-bar connections.
Therefore, high-level semantic information of music, such as chords (a fixed combination of adjacent pitches), is more likely to be captured by the n-Bar Level method.

\begin{table}
    \caption{Multi-level Object Selection Strategy.}
    \centering
    \resizebox{0.9\linewidth}{!}{
    \begin{tabular}{l|cc}
        \toprule
         & Element Level  & Token Level  \\
         \midrule
         Octuple Level & Octuple-Element Level  & Octuple-Token Level \\
         \midrule
         Bar Level & Bar-Element Level &  Bar-Token Level\\
         \midrule
         n-Bar Level & n-Bar-Element Level & n-Bar-Token Level\\
         \bottomrule
    \end{tabular}
    }
    \label{tab:table1}
\end{table}

\subsection{Pre-training PianoBART} \label{Pre-training PianoBART}
To train PinaoBART, we utilize the noising approaches of BART \cite{BART}, which consists of five transformations.
By combining these transformations, PinaoBART enables any type of corruption to the original music sequence, forcing the model to reason more about the musical context.
For each transformation, we apply different object selection methods.

\noindent
\textbf{(1) Token Masking.} 
Elements or tokens are randomly sampled and replaced with the [MASK] token.
For this task, we use four object selection methods, including the Octuple-Element Level, the Octuple-Token Level, the n-Bar-Element Level, and the n-Bar-Token Level.

\noindent
\textbf{(2) Token Deletion.} 
Some of the objects are randomly deleted with a probability of 15\%.
It is evident that deleting the element-level object would cause alignment issues. 
Therefore, we only allow the Octuple-Token Level and the n-Bar-Token Level selection methods for this task.

\noindent
\textbf{(3) Text Infilling.}
Spans of objects are replaced with a single [MASK] token. 
Similar to the Token Deletion task, we also employ the Octuple-Token Level and the n-Bar-Token Level selection methods for this task.

\noindent
\textbf{(4) Sentence Permutation.}
For text, sentence permutation refers to randomly shuffling the sentence order \cite{BART}. 
In music, however, there are few concepts corresponding to sentences. 
Obtaining such concepts from MIDI is not feasible, and manual annotation requires specialized knowledge.
To this end, we treat each bar as a ``simplified sentence''.
We adopt the Octuple-Token Level and the Bar-Token Level method to randomly split the music and shuffle.

\noindent
\textbf{(5) Document Rotations.}
A token is selected, and then the sequence is rotated around the chosen token. 
Since we only need to choose one token at a time, we only utilize the Octuple-Token Level choosing method for this task.

During pre-training, PinaoBART is optimized with a reconstruction loss—the cross-entropy between the decoder's output and the original token sequence.

\section{Experiments and Results} \label{Experiments and Results}
In this section, we first present the experimental setup in this study. 
Then we conduct a series of evaluations and analyze the results to verify the performance of PinaoBART.

\subsection{Experimental Setup} \label{Experimental Setup}
We pre-train PianoBART with 8 layers of 8 attention heads in each of the encoder and decoder, and a hidden size of 1024, resulting in 225M parameters.
The batch size is 16 sequences, each with a maximum length of 1024 octuple tokens. 
The training is conducted on two NVIDIA V100 GPUs for 3 days.
We use Adam optimizer and set the learning rate to 2e-5, and $L_2$ weight decay to 1e-2.
We clip the gradient with the maximum norm of 3.
The training is early stopped when the loss has not decreased for 30 consecutive epochs.
During pre-training, PianoBART adopts the proposed multi-level object selection strategy and randomly selects one of the 5 transformations to dynamically corrupt the data for each batch.

\subsection{Pre-training} \label{Pretraining}
We collect five public available piano MIDI datasets (Pop1K7 \cite{CP}, ASAP \cite{ASAP}, POP909 \cite{POP909}, Pianist8 \cite{MidiBERT}, and EMOPIA \cite{EMOPIA}) to train PianoBART.
These datasets contain Western classical music as well as piano covers of pop music, including 4166 pieces in total.
We convert MIDI files into Octuple sequences and split them into segments with 1024 tokens, which results in 8393 segments for pre-training.
MusicBERT \cite{MusicBERT} and MidiBERT \cite{MidiBERT} are used as baselines and are pre-trained with the same data and resources.
Each model was trained five times and the average performance is shown in Table \ref{tab:pretrain}.
Among the baselines, PianoBART achieves the shortest pre-training time with a speed faster than baselines, reaching the best reconstruction accuracy of over 96\%.

\subsection{Fine-tuning} \label{Fine-tuning}
We fine-tune PianoBART on two types of downstream tasks: music generation and understanding.
In this work, music generation involves conditioning the model with a fragment of piano performance as the prompt and producing a continuation. 
The encoder takes the prompt as input and the decoder autoregressively generates the target music.
We leverage the stochastic temperature-controlled sampling method \cite{CP} to improve the diversity of generated samples. 

\begin{table}
\caption{Pretraining Performance. 
}
\centering
\resizebox{0.95\linewidth}{!}{
\begin{tabular}{lcccc}
\hline
\textbf{Model} & \textbf{Time} & \textbf{Epoch} & \textbf{Parameter} & \textbf{Accuracy}\\ 
\hline
MusicBERT \cite{MusicBERT} & 10.06 d & 500 & 114 million & 76.01\% \\
MidiBERT \cite{MidiBERT} & 6.44 d & 500  & 111 million & 79.60\% \\ 
PianoBART & \textbf{3.19 d} & \textbf{268} & 225 million & \textbf{96.67\%} \\
\hline
\end{tabular}
}
\label{tab:pretrain}
\end{table}

\begin{table}
\caption{Results of music continuation on MAESTRO \cite{MAESTRO}.
}
\centering
\resizebox{\linewidth}{!}{
\begin{tabular}{lcccc}
\hline
\textbf{Model} & \textbf{$PFS_{GT} \uparrow$} & \textbf{$PFS_{prompt} \uparrow$} & \textbf{$PCHE \downarrow$} & \textbf{$GS \downarrow$}\\ 
\midrule
        Music Transformer \cite{MusicTransformer}    & 0.1721 & 0.1903 & 0.496 & 0.140\\
        Pop Music Transformer \cite{Pop_music_transformer}  & 0.7742 & 0.7647 & 0.360 & 0.015\\                       
        \midrule 
        PianoBART (w/o pre-train)   & 0.1502 & 0.1349 & 0.237 & 0.133\\
        PianoBART-simple            & \textbf{0.8495} & 0.8427 & 0.253 & 0.006\\
        PianoBART                   & 0.8245 & \textbf{0.8666} & \textbf{0.213} & \textbf{0.001}\\
        \bottomrule

\end{tabular}
}
\label{tab:Generation_MAESTRO}
\end{table}

The music understanding consists of two token-level tasks (velocity prediction and melody extraction), and two sequence-level tasks (emotion classification and composer classification). 
The same token sequence is fed into the encoder and decoder.
For token-level tasks, each input token corresponds to an output label.
The hidden state of the top-layer decoder is used to classify each token.
For the sequence-level tasks, the final hidden state of the decoder is fed into an additional Attention-based Weight Average Layer \cite{MidiBERT} to map the decoder’s output sequence to a single label.
We compare PianoBART with previous works on symbolic music understanding, including MusicBERT \cite{MusicBERT} and MidiBERT \cite{MidiBERT}.

\begin{table}
\caption{Results of music continuation on GiantMIDI \cite{GiantMidi}.
}
\centering
\resizebox{\linewidth}{!}{
\begin{tabular}{lcccc}
\hline
\textbf{Model} & \textbf{$PFS_{GT} \uparrow$} & \textbf{$PFS_{prompt} \uparrow$} & \textbf{$PCHE \downarrow$} & \textbf{$GS \downarrow$}\\ 
\midrule
        Music Transformer \cite{MusicTransformer}    & 0.1137 & 0.0867 & 0.476 & \textbf{0.018}\\
        Pop Music Transformer \cite{Pop_music_transformer}  & 0.5762 & 0.5640 & 0.378 & 0.019\\                       
        \midrule 
        PianoBART (w/o pre-train)   & 0.1793 & 0.1658 & \textbf{0.055} & 0.165\\
        PianoBART-simple            & 0.7334 & 0.6984 & 0.508 & 0.083\\
        PianoBART                   & \textbf{0.7708} & \textbf{0.7354} & 0.224 & 0.071\\
        \bottomrule

\end{tabular}
}
\label{tab:Generation_GiantMIDI}
\end{table}

\textbf{(1) Music Continuation.}
Since most music-generative models are auto-regressive sequence-to-sequence models \cite{MusicTransformer, Pop_music_transformer}, it's appropriate to evaluate with a continuation task.
We fine-tune PianoBART on two MIDI datasets: MAESTRO \cite{MAESTRO} includes 1276 recordings, and GiantMIDI-Piano \cite{GiantMidi} contains 7236 pieces of classical music, from which we select 1383 songs.
Given our resource constraints, we randomly crop segments with 1024 octuple tokens as the prompt and the following 1024 tokens as the continuation.

We compare PianoBART with Music Transformer \cite{MusicTransformer} and Pop Music Transformer \cite{Pop_music_transformer} on Pitch Fréchet Similarity (PFS), Pitch Class Histogram Entropy (PCHE), and Grooving pattern Similarity (GS).
PFS measures the pitch distance of generated results from the ground truth (GT) \cite{FD}. 
We also test the PFS between generated results and the given prompt to measure repeated patterns.
PCHE reflects the stability of the pitch distribution and GS measures the coherence of rhythm \cite{JazzTransformer}. 
Both PCHE and GS concern the music itself and we compute their absolute difference with GT. 
Results are shown in Table \ref{tab:Generation_GiantMIDI} and Table \ref{tab:Generation_MAESTRO}.
We observe that PianoBART outperforms baselines by a large margin, which shows the effectiveness of PianoBART.
We also see that PianoBART supports generating music of arbitrary length, while \cite{MusicTransformer, Pop_music_transformer} can only produce sequences of limited length.

\begin{table*}[thb]
\centering
\caption{The testing classification accuracy of different models on four music understanding task (velocity prediction, melody extraction, emotion classification and composer classification).
The best accuracy is achieved by PianoBART in all these tasks.
}
\resizebox{0.8\textwidth}{!}{
    \begin{tabular}{lccccc}
    \toprule
    {} & \multicolumn{2}{c}{\textbf{Token-level Tasks}} & \multicolumn{3}{c}{\textbf{Sequence-level Tasks}}  \\
    \cmidrule[0.5pt](rl){2-3}
    \cmidrule[0.5pt](rl){4-6}
    
        \textbf{Model} & \textbf{Velocity} & \textbf{Melody} & \textbf{Emotion} & \textbf{Composer (Pianoist8)}  & \textbf{Composer (ASAP)}\\ 
        \midrule
        MusicBERT \cite{MusicBERT} & 51.23\% & 92.47\% & 71.06\% & 86.05\% & 94.27\% \\
        MidiBERT \cite{MidiBERT} & 48.57\% & 92.53\% & 67.59\% & 79.07\% & 96.18\%\\ 
        \midrule 
        PianoBART (w/o pre-training) & 38.55\% & 82.40\% & 58.33\% & 69.77\% & 78.34\%\\
        PianoBART-simple & 51.57\% & 92.50\% & 66.67\% & 83.72\% & 96.32\%\\
        PianoBART & \textbf{51.63\%} & \textbf{92.62\%} & \textbf{73.15\%} & \textbf{88.37\%} & \textbf{97.45\%}\\
        \bottomrule
    \end{tabular}
}
\label{tab:understanding}
\end{table*}

\textbf{(2) Velocity Prediction.}
The velocity in music indicates the level of dynamics and corresponds to the perceptual loudness of notes.
Learning the velocity is helpful in modeling expressive piano performance \cite{VelocityPrediction_1, VelocityPrediction_2}.
Following \cite{MidiBERT}, in our experiment, we quantize MIDI velocity values (0-127) into 6 levels. 
Velocity prediction is regarded as a 6-class classification task.
We adopt the GiantMidi dataset \cite{GiantMidi} and train the model to predict the velocity level for each note.
Table \ref{tab:understanding} shows the classification accuracy.
It's apparent that the performance is generally not high, which may be because velocity is a rather subjective factor related to the dynamics of performers.

\textbf{(3) Melody Extraction.} 
Melody is the most intuitive and important element of a musical composition.
In the symbolic domain, melody extraction is to identify the melody notes in a MIDI file of polyphonic piano music \cite{MelodyExtraction_1, MelodyExtraction_2}, which is a critical token-level understanding task.
We fine-tune PianoBART on the POP909 dataset \cite{POP909} that contains piano covers of 909 pop songs, with labels of melody, bridge, and accompaniment for each note.
The performance is evaluated by classification accuracy and the results are shown in Table \ref{tab:understanding}.
There is little performance difference between the compared methods, with PianoBART achieving slightly better accuracy.

\textbf{(4) Emotion Recognition.}
Music is a natural carrier to express and convey emotion.
Understanding the overall emotion of music \cite{MusiceEmotionRecognition, MusicMoodClassification} is of great significance for topics such as music recommendation and personalized music generation. 
In this paper, emotion recognition is treated as a 4-class classification problem.
We adopt the EMOPIA \cite{EMOPIA}, an emotion-labeled symbolic music dataset for this task.
The emotional annotation of each clip is labeled using the 4-class taxonomy (HVHA, HVLA, LVHA, LVLA). 
Table \ref{tab:understanding} shows that PianoBART outperforms compared baselines, showing its ability in symbolic-domain emotion recognition.

\textbf{(5) Composer Classification.} 
Recognizing a composer of a piece \cite{ComposerClassification_1, ComposerClassification_2} used to be reserved for experts in music theory, which is a fine-grained discriminative task compared to genre or style classification. 
We use the Pianist8 dataset \cite{MidiBERT} and the ASAP dataset \cite{ASAP} for this task.
Pianist8 consists of 411 original piano pieces performed by eight composers.
ASAP contains 1068 MIDI pieces of Western classical piano music from 15 composers.
Table \ref{tab:understanding} demonstrates the strengths of PianoBART on this professional sequence-level music comprehension task.

\begin{figure}[thb] 
\centering
  \includegraphics[width=0.43\textwidth]{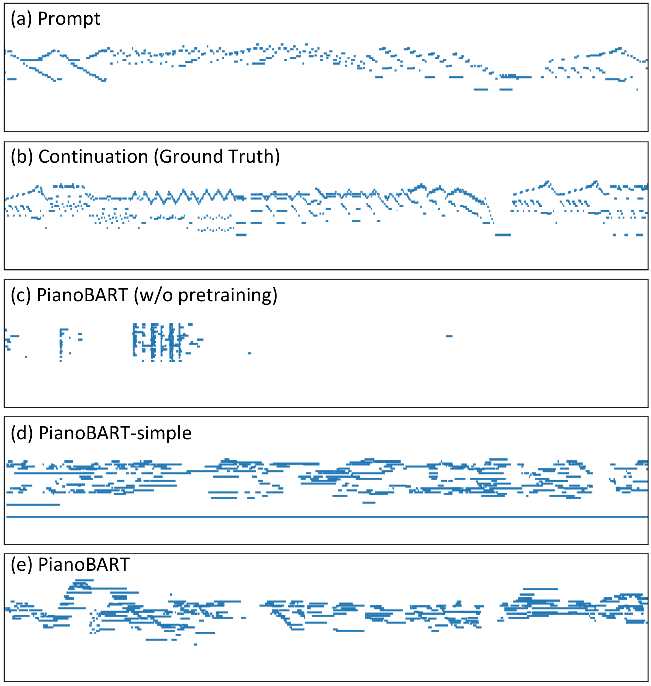}
  \caption{
Visualization results of generated examples on ablation variants.
PianoBART (w/o pretraining), PianoBART-simple, and PianoBART are all continued from Prompt.
} 
\label{fig:Ablation}
\end{figure}

\subsection{Ablation Study} \label{Ablation Study}
We design two variants (PianoBART (w/o pretraining) and PianoBART-simple) to validate the effects of pre-training and the proposed multi-level object selection strategy.
(1) PianoBART (w/o pretraining) does not initialize the model with the pre-trained parameters, and only uses the data of downstream tasks to train PianoBART from scratch.
(2) PianoBART-simple only uses the Octuple-Token Level method to select the object to be corrupted for any pre-training transformation.

Table \ref{tab:Generation_GiantMIDI} and Table \ref{tab:Generation_MAESTRO} reflect the results of ablated variants on music continuation.
Figure \ref{fig:Ablation} shows the piano-roll visualization of MIDI produced by ablated variants.
We can see that the lack of pre-training performs worse on PFS and struggles in continuous music generation.
PianoBART-simple achieves better PFS scores but lacks clear and coherent musical patterns.
Notably, PianoBART generates harmonious long-term music with coherent rhythm, showing the effectiveness of pre-training and the proposed multi-level selection strategy.
We provide MIDI demos in the supplementary material and recommend readers to listen for an intuitive experience.

Table \ref{tab:understanding} shows the ablation results on four music understanding tasks.
It's obvious that the absence of pre-training significantly affects the accuracy, while using the pre-trained model to initialize and fine-tune can improve the performance.
The results can be further improved by using PianoBART's multi-level object selection strategy, which demonstrates the effectiveness of our method.

\section{Conclusion}\label{Conclusion}
In this paper, we propose PianoBART, a comprehensive pre-trained model designed for symbolic music understanding and generation. 
By introducing the BART framework and devising a multi-level object selection strategy, PianoBART exhibits remarkable performance in generating coherent music and understanding musical patterns. 
The ablation results demonstrate the effectiveness of the pre-training and the proposed multi-level object selection strategy.
PianoBART holds significant potential for advancing music study and creation. 
Future works involve further enhancements in the model’s performance and the incorporation of expert knowledge.

\section*{Acknowledgments}
This work was supported by the Guandong Basic and Applied Basic Research Foundation (Grant No.2022A1515011425).

\bibliographystyle{IEEEtran}
\bibliography{IEEEabrv,icme2023template}

\begin{thebibliography}{10}
\providecommand{\url}[1]{#1}
\csname url@samestyle\endcsname
\providecommand{\newblock}{\relax}
\providecommand{\bibinfo}[2]{#2}
\providecommand{\BIBentrySTDinterwordspacing}{\spaceskip=0pt\relax}
\providecommand{\BIBentryALTinterwordstretchfactor}{4}
\providecommand{\BIBentryALTinterwordspacing}{\spaceskip=\fontdimen2\font plus
\BIBentryALTinterwordstretchfactor\fontdimen3\font minus \fontdimen4\font\relax}
\providecommand{\BIBforeignlanguage}[2]{{%
\expandafter\ifx\csname l@#1\endcsname\relax
\typeout{** WARNING: IEEEtran.bst: No hyphenation pattern has been}%
\typeout{** loaded for the language `#1'. Using the pattern for}%
\typeout{** the default language instead.}%
\else
\language=\csname l@#1\endcsname
\fi
#2}}
\providecommand{\BIBdecl}{\relax}
\BIBdecl

\bibitem{MusicBERT}
M.~Zeng, X.~Tan, R.~Wang, Z.~Ju, T.~Qin, and T.-Y. Liu, ``Musicbert: Symbolic music understanding with large-scale pre-training,'' in \emph{Findings of the Association for Computational Linguistics: ACL-IJCNLP 2021}, 2021, pp. 791--800.

\bibitem{MidiBERT}
Y.-H. Chou, I.~Chen, C.-J. Chang, J.~Ching, Y.-H. Yang \emph{et~al.}, ``Midibert-piano: large-scale pre-training for symbolic music understanding,'' \emph{arXiv preprint arXiv:2107.05223}, 2021.

\bibitem{MusicTransformer}
C.-Z.~A. Huang, A.~Vaswani, J.~Uszkoreit, I.~Simon, C.~Hawthorne, N.~Shazeer, A.~M. Dai, M.~D. Hoffman, M.~Dinculescu, and D.~Eck, ``Music transformer: Generating music with long-term structure,'' in \emph{International Conference on Learning Representations}, 2018.

\bibitem{StyleMusicGeneration}
Y.-Q. Lim, C.~S. Chan, and F.~Y. Loo, ``Style-conditioned music generation,'' in \emph{2020 IEEE International Conference on Multimedia and Expo (ICME)}.\hskip 1em plus 0.5em minus 0.4em\relax IEEE, 2020, pp. 1--6.

\bibitem{Pop_music_transformer}
Y.-S. Huang and Y.-H. Yang, ``Pop music transformer: Beat-based modeling and generation of expressive pop piano compositions,'' in \emph{Proceedings of the 28th ACM international conference on multimedia}, 2020, pp. 1180--1188.

\bibitem{InformationLeakage}
Z.~Shen, L.~Yang, Z.~Yang, and H.~Lin, ``More than simply masking: Exploring pre-training strategies for symbolic music understanding,'' in \emph{Proceedings of the 2023 ACM International Conference on Multimedia Retrieval}, 2023, pp. 540--544.

\bibitem{CP}
W.-Y. Hsiao, J.-Y. Liu, Y.-C. Yeh, and Y.-H. Yang, ``Compound word transformer: Learning to compose full-song music over dynamic directed hypergraphs,'' in \emph{Proceedings of the AAAI Conference on Artificial Intelligence}, vol.~35, no.~1, 2021, pp. 178--186.

\bibitem{CSIBERT}
Z.~Zhao, T.~Chen, F.~Meng, H.~Li, X.~Li, and G.~Zhu, ``Finding the missing data: A bert-inspired approach against package loss in wireless sensing,'' \emph{arXiv preprint arXiv:2403.12400}, 2024.

\bibitem{CV_mask}
Y.~Li, H.~Fan, R.~Hu, C.~Feichtenhofer, and K.~He, ``Scaling language-image pre-training via masking,'' in \emph{Proceedings of the IEEE/CVF Conference on Computer Vision and Pattern Recognition}, 2023, pp. 23\,390--23\,400.

\bibitem{EL_pretrain}
Z.~Lyu, Y.~Li, G.~Zhu, J.~Xu, H.~V. Poor, and S.~Cui, ``Rethinking resource management in edge learning: A joint pre-training and fine-tuning design paradigm,'' \emph{arXiv preprint arXiv:2404.00836}, 2024.

\bibitem{BERT}
J.~D. M.-W.~C. Kenton and L.~K. Toutanova, ``Bert: Pre-training of deep bidirectional transformers for language understanding,'' in \emph{Proceedings of naacL-HLT}, vol.~1, 2019, p.~2.

\bibitem{Transformer}
A.~Vaswani, N.~Shazeer, N.~Parmar, J.~Uszkoreit, L.~Jones, A.~N. Gomez, {\L}.~Kaiser, and I.~Polosukhin, ``Attention is all you need,'' \emph{Advances in neural information processing systems}, vol.~30, 2017.

\bibitem{Popmag}
Y.~Ren, J.~He, X.~Tan, T.~Qin, Z.~Zhao, and T.-Y. Liu, ``Popmag: Pop music accompaniment generation,'' in \emph{Proceedings of the 28th ACM international conference on multimedia}, 2020, pp. 1198--1206.

\bibitem{BART}
M.~Lewis, Y.~Liu, N.~Goyal, M.~Ghazvininejad, A.~Mohamed, O.~Levy, V.~Stoyanov, and L.~Zettlemoyer, ``Bart: Denoising sequence-to-sequence pre-training for natural language generation, translation, and comprehension,'' in \emph{Proceedings of the 58th Annual Meeting of the Association for Computational Linguistics}, 2020, pp. 7871--7880.

\bibitem{POP909}
Z.~Wang, K.~Chen, J.~Jiang, Y.~Zhang, M.~Xu, S.~Dai, X.~Gu, and G.~Xia, ``Pop909: A pop-song dataset for music arrangement generation,'' \emph{arXiv preprint arXiv:2008.07142}, 2020.

\bibitem{ASAP}
F.~Foscarin, A.~Mcleod, P.~Rigaux, F.~Jacquemard, and M.~Sakai, ``Asap: a dataset of aligned scores and performances for piano transcription,'' in \emph{International Society for Music Information Retrieval Conference}, no. CONF, 2020, pp. 534--541.

\bibitem{EMOPIA}
H.-T. Hung, J.~Ching, S.~Doh, N.~Kim, J.~Nam, and Y.-H. Yang, ``Emopia: A multi-modal pop piano dataset for emotion recognition and emotion-based music generation,'' in \emph{International Society for Music Information Retrieval Conference, ISMIR 2021}.\hskip 1em plus 0.5em minus 0.4em\relax International Society for Music Information Retrieval, 2021.

\bibitem{Roberta}
Y.~Liu, M.~Ott, N.~Goyal, J.~Du, M.~Joshi, D.~Chen, O.~Levy, M.~Lewis, L.~Zettlemoyer, and V.~Stoyanov, ``Roberta: A robustly optimized bert pretraining approach,'' \emph{arXiv preprint arXiv:1907.11692}, 2019.

\bibitem{MAESTRO}
C.~Hawthorne, A.~Stasyuk, A.~Roberts, I.~Simon, C.-Z.~A. Huang, S.~Dieleman, E.~Elsen, J.~Engel, and D.~Eck, ``Enabling factorized piano music modeling and generation with the {MAESTRO} dataset,'' in \emph{International Conference on Learning Representations}, 2019.

\bibitem{GiantMidi}
Q.~Kong, B.~Li, J.~Chen, and Y.~Wang, ``Giantmidi-piano: A large-scale midi dataset for classical piano music,'' 2022.

\bibitem{FD}
T.~Eiter and H.~Mannila, ``Computing discrete fr{\'e}chet distance,'' 1994.

\bibitem{JazzTransformer}
S.-L. Wu and Y.-H. Yang, ``The jazz transformer on the front line: Exploring the shortcomings of ai-composed music through quantitative measures,'' \emph{arXiv preprint arXiv:2008.01307}, 2020.

\bibitem{VelocityPrediction_1}
D.~Jeong, T.~Kwon, Y.~Kim, K.~Lee, and J.~Nam, ``Virtuosonet: A hierarchical rnn-based system for modeling expressive piano performance.'' in \emph{ISMIR}, 2019, pp. 908--915.

\bibitem{VelocityPrediction_2}
D.~Jeong, T.~Kwon, Y.~Kim, and J.~Nam, ``Graph neural network for music score data and modeling expressive piano performance,'' in \emph{International conference on machine learning}.\hskip 1em plus 0.5em minus 0.4em\relax PMLR, 2019, pp. 3060--3070.

\bibitem{MelodyExtraction_1}
F.~Simonetta, C.~Cancino-Chac{\'o}n, S.~Ntalampiras, and G.~Widmer, ``A convolutional approach to melody line identification in symbolic scores,'' \emph{arXiv preprint arXiv:1906.10547}, 2019.

\bibitem{MelodyExtraction_2}
W.~Chai and B.~Vercoe, ``Melody retrieval on the web,'' in \emph{Multimedia Computing and Networking 2002}, vol. 4673.\hskip 1em plus 0.5em minus 0.4em\relax SPIE, 2001, pp. 226--241.

\bibitem{MusiceEmotionRecognition}
J.~Zhao, G.~Ru, Y.~Yu, Y.~Wu, D.~Li, and W.~Li, ``Multimodal music emotion recognition with hierarchical cross-modal attention network,'' in \emph{2022 IEEE International Conference on Multimedia and Expo (ICME)}.\hskip 1em plus 0.5em minus 0.4em\relax IEEE, 2022, pp. 1--6.

\bibitem{MusicMoodClassification}
Y.~Xiong, F.~Su, and Q.~Wang, ``Automatic music mood classification by learning cross-media relevance between audio and lyrics,'' in \emph{2017 IEEE international conference on multimedia and expo (ICME)}.\hskip 1em plus 0.5em minus 0.4em\relax IEEE, 2017, pp. 961--966.

\bibitem{ComposerClassification_1}
T.~Tsai and K.~Ji, ``Composer style classification of piano sheet music images using language model pretraining,'' \emph{arXiv preprint arXiv:2007.14587}, 2020.

\bibitem{ComposerClassification_2}
Q.~Kong, K.~Choi, and Y.~Wang, ``Large-scale midi-based composer classification,'' \emph{arXiv preprint arXiv:2010.14805}, 2020.

\end{thebibliography}







\end{document}